\documentclass[10pt,letterpaper,twocolumn]{article}
\usepackage{ol2}

\usepackage{graphicx}
\def\sech{\mathop{\rm sech}\nolimits}

\begin{document}
\twocolumn[ %% activate for two-column option
\title{Vector vortex solitons in nematic liquid crystals}

\author{Zhiyong Xu$^1$$^{,*}$, Noel F. Smyth$^2$, Antonmaria A. Minzoni$^3$, and Yuri S. Kivshar$^1$}

\address{$^1$Nonlinear Physics Center, Research School of Physics
and Engineering,\\ Australian National University, Canberra ACT 0200,
Australia\\
$^2$School of Mathematics and Maxwell Institute for
Mathematical Sciences,\\University of Edinburgh,
Edinburgh, Scotland EH9 3JZ, U.K.\\
$^3$Fenomenos Nonlineales y Mec\'anica (FENOMEC), Department of
Mathematics and Mechanics, Instituto de Investigaci\'on en
Matem\'aticas Aplicadas y Sistemas, Universidad Nacional
Aut\'onoma de M\'exico, 01000 M\'exico D.F., M\'exico\\
$^*$Corresponding author: xzy124@rsphysse.anu.edu.au}

\begin{abstract}
We analyze the existence and stability of two-component vector solitons in nematic liquid
crystals for which one of the components carries angular momentum and describes a vortex beam.
We demonstrate that the nonlocal, nonlinear response can dramatically enhance the field coupling
leading to the stabilization of the vortex beam when the amplitude of the second beam exceeds some
threshold value.  We develop a variational approach to describe this effect analytically.
\end{abstract}

] %% activate for two-column option

\maketitle

Optical vortices are usually introduced as phase singularities in
diffracting optical beams~\cite{review_soskin} and can be
generated in both linear and nonlinear media.  The well known
effect accompanying the propagation of such singular beams and vortex solitons
in self-focusing, nonlinear media is vortex breakup into several
fundamental solitons via a symmetry-breaking azimuthal instability~\cite{review_our}.
However, recent numerical studies have revealed that spatially localized
vortex solitons can be stabilized in highly nonlocal self-focusing
nonlinear media~\cite{kivshar,wieslaw,oe_anton}.
This stabilization effect was later explained analytically~\cite{pre_our} by employing
a modulation theory for the vortex parameters based on an averaged Lagrangian.

Spatial optical vector solitons can form when several beams
propagate together, interacting parametrically or via the effect of
cross-phase modulation~\cite{book}.  The simplest vector solitons are known
as shape-preserving, self-localized solutions of coupled nonlinear
evolution equations~\cite{book}.  A novel class of vector solitons
in the form of two color spatial solitons in a highly
nonlocal and anisotropic Kerr-like medium were predicted
to exist in nematic liquid crystals~\cite{pre_xu,Yaroslav1,prl_gaetano}.
The first experimental observations of anisotropic, nonlocal vector solitons
in unbiased nematic liquid crystals were reported by Alberucci
{\em et al.}~\cite{prl_gaetano}, who investigated the interaction between two beams
of different wavelengths and observed that two extraordinarily polarized beams
of different wavelengths can nonlinearly couple, compensating for the beam
walk-off, so creating a vector soliton.

The main purpose of this Letter is twofold.  Firstly, we introduce
a novel class of vector solitons in nonlocal, nonlinear media, such as
nematic liquid crystals and study their properties.  These vector solitons appear
as two color, self-trapped beams for which one of the components carries angular momentum and
describes a vortex beam.  Secondly, we demonstrate that the nonlocal, nonlinear response may
dramatically enhance the field coupling, leading to the stabilization of the vortex for much
weaker nonlocality when the amplitude of the second beam exceeds some threshold value.
We develop a variational approach to describe this effect analytically.

We consider the propagation of two light beams of different wavelengths in a cell
filled with a nematic liquid crystal. The light propagates in the $z$ direction, with the $(x,y)$
plane orthogonal to this.  The electric fields of the light beams are assumed to be polarized in
the $x$ direction.  The system of coupled equations for the dimensionless complex
field amplitudes $u$ and $v$ can be written in the form
\begin{eqnarray}
\label{eq:model}
&&i\frac{\partial u}{\partial z}+\frac{1}{2}\left(\frac{\partial^{2}u}{\partial x^{2}}+\frac{\partial^{2}u}{\partial y^{2}}\right)+ 2u \theta=0, \nonumber \\
&&i\frac{\partial v}{\partial z}+\frac{1}{2}\left(\frac{\partial^{2}v}{\partial x^{2}}+\frac{\partial^{2}v}{\partial y^{2}}\right)+ 2v \theta=0, \nonumber \\
&&\nu \left(\frac{\partial^{2}\theta}{\partial x^{2}}+\frac{\partial^{2}\theta}{\partial y^{2}}\right)- 2\theta= -2(|u|^{2}+|v|^{2}),
\end{eqnarray}
where $\theta$ describes the change of the director angle from the pre-tilt state, which is related to the nonlinear correction to the optical refractive index.  In Eqs.~(\ref{eq:model}) the longitudinal ($z$) and transverse ($x,y$) coordinates are normalized to the diffraction length and the beam width, respectively. The parameter $\nu$ describes the degree of nonlocality of the nonlinear response.  When $\nu\rightarrow 0$, Eqs.~(\ref{eq:model}) reduce to the Manakov vector nonlinear equations.  The system (\ref{eq:model}) conserves the energy flow $P=P_{1}+P_{2}=\int\int^{+\infty}_{-\infty}(|u|^{2}+|v|^{2})dxdy$.

We are interested in a special class of vector solitons for which one of the components carries angular momentum and the other component describes a spatially localized mode in the form of a spatial bright
beam.  Solutions of this type have been discussed earlier for nonlinear systems with a local response~\cite{vec_1,vec_2} and they have been shown to be unstable in a large region of their
existence domain~\cite{vec_3}.  For our system described by Eqs.\ (\ref{eq:model}) such solutions can be
found in the form $u=w_{1}(r)\texttt{exp}(ib_{1}z)$ and $v=w_{2}(r)\texttt{exp}(i\phi)\texttt{exp}(ib_{2}z)$, where $w_{1}(r)$ and $w_{2}(r)$ are real
functions describing the beam envelopes, $b_{1,2}$ are real propagation constants and $r=\sqrt{x^{2}+y^{2}}$ in the radial coordinate.  The resulting system of equations obtained
after substitution of these solution forms into Eqs.\ (\ref{eq:model}) is solved using a standard
numerical relaxation method.  Without loss of generality, we search for solutions with
$b_{2} \leq b_{1}$ and set $b_{1}=3$ to investigate the properties of vector vortex solitons
by varying the propagation constant $b_{2}$ and the nonlocality parameter $\nu$.

%%%%%%%%%%%%%%%%%%%%%%%%%%%%%%%%%%%%

\begin{figure}[htb]
\centerline{\includegraphics[width=6.8cm]{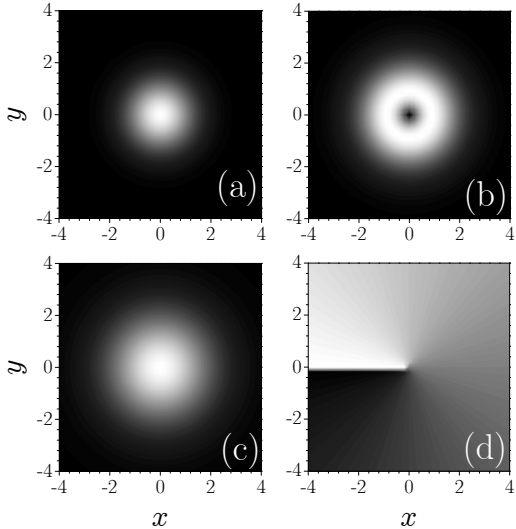}}
\caption{(a,b) Field distribution for (a) bright soliton and (b) vortex beam. (c) Nonlinear correction to
the refractive index. (d) Phase structure of the corresponding vortex beam shown in (b).
Here $\nu=1$, $b_{1}=3$, and $b_{2}=1.8$.}
\label{fig1}
\end{figure}

\begin{figure}[htb]
\centerline{\includegraphics[width=7.5cm]{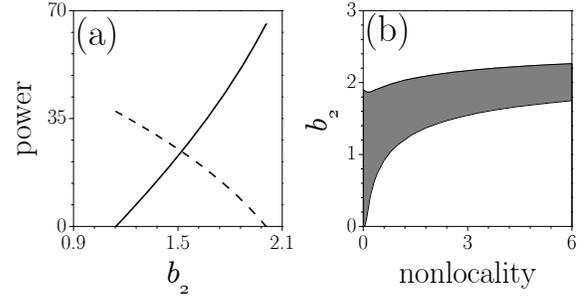}} \caption{(a) Power of bright (dashed) and
vortex (solid) beams for $\nu=1$. (b) Existence domain (gray) of vector solitons as a function of
nonlocality parameter $\nu$ (at $b_{1}=3$).} \label{fig2}
\end{figure}

Figure~\ref{fig1} presents an example of vector vortex solitons for Eqs.\ (\ref{eq:model}) for which
one component has the shape of a bright soliton [Fig.~\ref{fig1}(a)] and the other component carries
angular moment, so forming a vortex soliton [Fig.~\ref{fig1}(b,d)].  Due to the physical nature of the
nonlocal response of the nematic liquid crystal, we notice that the refractive index
change features a bell-shaped distribution [Fig.~\ref{fig1}(c)], even though there is a singularity
in the center of the vortex beam, this being crucial for the stabilization of vortex solitons.  As shown
in Fig.~\ref{fig2}(a), for fixed propagation constant $b_{1}$ and nonlocality parameter $\nu$ the
power of the vortex beam is a monotonically increasing function of the propagation constant $b_{2}$,
while the power of the bright soliton decreases monotonically.  It is important to note that vector
vortex solitons exist in a finite band of the propagation constant $b_{2}$.  At the lower band edge
the vortex beam vanishes and one obtains a scalar bright soliton.  However, at the upper band edge the
beam with a bell shape vanishes, so that the vector vortex soliton transforms into a scalar vortex soliton.
We find that the existence domain of vector vortex solitons shrinks with increasing nonlocality parameter
$\nu$ [Fig.~\ref{fig2}(b)].

\begin{figure}[htb]
\centerline{\includegraphics[width=4.8cm]{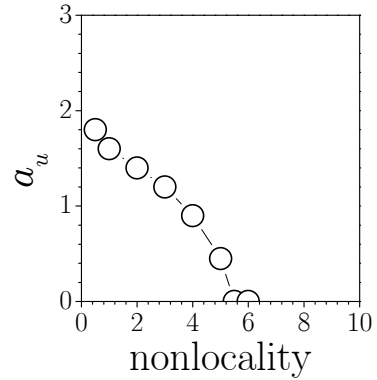}} \caption{Dependence of
the critical value of the amplitude for a bright soliton on the nonlocality degree $\nu$.} \label{fig3}
\end{figure}

\begin{figure}[htb]
\centerline{\includegraphics[width=7cm]{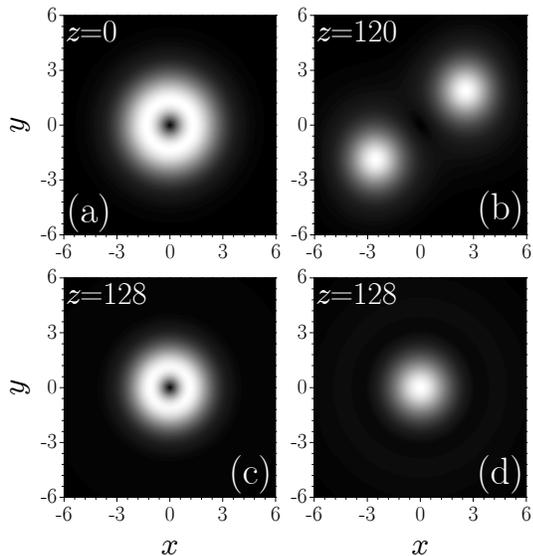}} \caption{(a,b) Input and output
of the intensity for an unstable vortex beam propagating alone in the medium. (c,d) Stable propagation
of the vortex beam coupled to the bright component.  Here $\nu=4$ and $a_{u}^{T}=0.9$.} \label{fig4}
\end{figure}

As one of the central results we find that a bright beam with a finite amplitude can stabilize
an otherwise unstable vortex beam.  To address this issue we performed extensive numerical
simulations of Eqs.\ (\ref{eq:model}) using the beam propagation method.  Firstly, we employed a
stationary form of the vortex beam (in order to minimize radiation) as an input beam for the
$v$ component only (namely, there is no bright soliton as an input for the $u$ component),
noting that a vortex beam is unstable when propagating alone.  Then we added a Gaussian pulse
to the $u$ component and studied the dynamics of a vortex beam propagating together with a Gaussian
beam by varying the nonlocality parameter $\nu$.  Our main results are presented in Fig.~\ref{fig3},
from which one can see that for low nonlocality, a higher amplitude of the bright beam is required
for stabilization of the vortex beam, while for high enough nonlocality, the vortex beam is
observed to be stable, even when propagating alone.  Figure~\ref{fig4} shows some illustrative examples.
It is clearly seen that when the vortex beam propagates alone it becomes unstable and breaks up
into two filaments [see Fig.~\ref{fig4}(b)].  However, when we add a Gaussian beam with amplitude
0.9 above the threshold (namely, $a_{u}^{T}=0.9$), then the vortex beam co-propagates with the
Gaussian beam in a stable manner.  Thus, we draw the conclusion that a nonlocal, nonlinear response can
dramatically enhance the field coupling, leading to the stabilization of the vortex soliton when
the amplitude of the Gaussian beam exceeds some threshold value.

Let us now give a sketch of the modulation theory analysis that allows deeper insight into
the effect of the vortex stabilization and which will explain in a quantitative manner the
results shown in Fig.~\ref{fig3}.  Using the two-component
equations re-written in a Lagrangian formulation \cite{pra_assanto}, we employ
the trial functions for the vortex, its soliton component and the director angle,
\begin{eqnarray}
\label{eq:trial}
&& u = a_{u} \sech (r/w_{u}) \: e^{i\sigma_{u} z}, \nonumber \\
&& v = a_{v} r e^{-r/w_{v}} e^{i\phi + i\sigma_{v} z} + ig e^{i\phi +
 i\sigma_{v} z}, \nonumber \\
&&\theta = \alpha_{u} \sech^{2} (r/\beta_{u}).
\end{eqnarray}
These functions are substituted into the Lagrangian, which is then averaged
by integrating in $r$ and $\phi$ over the plane.  This procedure gives the
averaged Lagrangian ${\cal L} = {\cal L}_{u} + {\cal L}_{v} + {\cal L}_{uv}$,
where ${\cal L}_{u}$ is given in Ref.~\cite{josab_minzoni} and ${\cal L}_{v}$ is given
in Ref.~\cite{pre_our}.  The important interaction Lagrangian between the vortex and the
soliton is $ {\cal L}_{uv} = (a_{u}^{2}w_{u}^{2})(2\sqrt{2q\nu})^{-1}\: a^{2}_{v}w_{v}^{2}$.

The results of Minzoni {\em et al} \cite{pre_our} show that a vortex is
stabilized as its width decreases and its amplitude increases.  Therefore, we
just need to show that the width of the vortex decreases as the amplitude of
the soliton in the other component increases.  If $A_{v} = a_{v}w_{v}e^{-1}$
is the amplitude of the vortex, then from Ref.~\cite{PRA_minzoni}
\begin{equation}
 \frac{e^{2}A_{v}^{2}}{8\sqrt{\nu}} \: w_{v}^{2} + (a_{u}w_{u})^{2} w_{v} - \frac{3}{2} = 0.
 \label{e:ampwidthv}
\end{equation}
Using the vortex width determined by this expression in the stability threshold of
Ref.~\cite{pre_our}, we find that the vortex is stable provided
\begin{equation}
 \frac{405}{128 \nu} \: A_{v}^{2}w_{v}^{4} < 14.4.
 \label{e:stab}
\end{equation}
Combining this criterion with the amplitude-width relation (\ref{e:ampwidthv}), it
can be seen that the vortex stabilizes for lower values of nonlocality parameter $\nu$ as the
amplitude $a_{u}$ of the bright soliton increases, which explains the stability results
shown in Fig.~\ref{fig3} and Fig.~\ref{fig4}.

In conclusion, we have described theoretically a novel type of stable vector vortex soliton in nonlocal, nonlinear media, such as nematic liquid crystals.  These solitons appear in the form of two color
self-trapped beams for which one of the components carries angular momentum and is stabilized by the nonlocality enhanced beam coupling with the other spatially localized beam.  We have studied the effect of this stabilization numerically and have also developed a variational approach to describe it analytically.

This research was supported by the Australian Research Council and the Engineering and
Physical Sciences Research Council (EPSRC) under Grant No.\ EP/D075947/1.  The authors thank
G. Assanto and A. Desyatnikov for useful discussions.

\end{document}